\documentclass[aps,amsfonts,superscriptaddress, nofootinbib, floatfix, 10pt,twocolumn]{revtex4-1}
\usepackage{epsfig}
\usepackage{graphicx}
\usepackage{amsmath}
\usepackage[latin1]{inputenc}
\usepackage{amsbsy}
\usepackage{color}
\usepackage{epsfig}
\usepackage{graphicx}
\usepackage{amsmath}
\usepackage{amssymb}
\usepackage{bbm}
\usepackage{amsbsy}
\usepackage{slashed}
\DeclareMathOperator{\tr}{tr}

\usepackage{xcolor}

\begin{document}
\title{Nonlocal Nambu-Jona-Lasinio model with a fractional Lorentzian regulator in the real time formalism.}
\author{M. Loewe}
\affiliation{Facultad de F\'isica, Pontificia Universidad Cat\'olica de Chile, Casilla 306, Santiago 22, Chile.}
\affiliation{Centre for Theoretical Physics and Mathematical Physics, University of Cape Town, Rondebosch 7700, South Africa.}
\author{F. Marquez}
\affiliation{Facultad de F\'isica, Pontificia Universidad Cat\'olica de Chile, Casilla 306, Santiago 22, Chile.}
\author{C. Villavicencio}
\affiliation{Instituto de Ciencias B\'asicas, Universidad Diego Portales, 
Casilla 298-V, Santiago, Chile.}

\begin{abstract}
In this article we study the finite temperature and chemical potential effects in a nonlocal Nambu-Jona-Lasinio model in the real time formalism. We make the usual Wick rotation to get from imaginary to real time formalism. In doing so, we need to define our regulator in the complex plane $q^2$. This definition will be crucial in our later analysis. We study the poles in the propagator of this model and conclude that only some of them are of interst to us. Once we have a well defined model in real time formalism, we look at the chiral condensate to find the temperature at which chiral symmetry restoration will occur. We find a second order phase transition that turns to a first order one for high enough values of the chemical potential.
\end{abstract}

\maketitle

\section{Introduction}

The study of QCD in the nonperturbative regime is a highly interesting topic. Important features of QCD, such as confinement or the QCD phase diagram, cannot be described through a perturbative analysis of the theory. Because of this, several methods have been developed in order to deal with the nonperturbative sector, such as lattice QCD. However succesful, lattice QCD does not seem to be an appropiate tool for studying problems with finite baryon chemical potential because of the well known ``sign problem'' \cite{Lattice1, Lattice2}. Another method frequently used to study the nonperturbative sector of QCD, is the use of effective models such as the Nambu-Jona-Lasinio (NJL) model. This was originally developed as a model of interacting nucleons \cite{Jona, Lasinio}, however, nowadays it is vastly used as a model of interacting quarks to explore finite temperature and density effects \cite{Loewe1, Buballa, Klevansky, Blaschke1, Blaschke2}.\\

The nonlocal NJL (nNJL) model is a generalization of the NJL model \cite{nNJL1,nNJL2}. The model has a nonlocal interaction modulated by a regulator. This regulator can take a variety of forms, inspired by different models \cite{Scoccola6, Michal1}.\\

In the last years, some regulators were proposed in order to reproduce lattice simulations of the light quarks propagator  \cite{Scoccola1}. Two regulators are associated with the renormalization function and self-energy. These kind of regulators are interesting due to their analytic structure. They exhibit a cut in the complex plane, which is one of the features we are interested in treating the system in a real time formalism \cite{Loewe2}. We want to study the effects of these kind of regulators in the presence of temperature and density effects.\\

Temperature and chemical potential effects are usually introduced in the nNJL model through the Matsubara (imaginary time) formalism \cite{LeBellac, Das}. 
However, in this case the sums of Matsubara frequencies are an issue because of 
the complicated shape the regulators may have. A real time formalism was 
initially proposed to avoid the necesity of truncating the Matsubara frequencies 
in numerical calculations.
This formalism provides a description of the effective quarks  which we find to be quite insightful, since it provides us with a clear intepretation of confinement effects 
\cite{nNJL1, nNJL2}. 
The main purpose behind working with the real time formalism in such kind of effective models is to achieve a description of the system in terms of quasiparticles. 
The resulting quasiparticles will be expressed in terms of a mass and a decay width, allowing us to understand which of them will be relevant for the description of the system, and which not.
Those too massive will be not accesible, and particles with a big decay width are too unstable.
This is the case indeed when dealing
with the high temperature regime, with a Lorentzian
regulator, where only a few number of poles contribute to
the dynamics of the system, unlike the Gaussian regulator
case \cite{Loewe2, Buballa2}.
Real time formalism is also interesting since it allows us 
to study phenomena beyond thermodynamic equilibrium through the 
Schwinger-Keldysh formalism \cite{Keldysh1, Keldysh2, Keldysh3}.  
The use of a Lorentzian regulator has the advantage of generating a finite 
number of poles, contrary to the case of the Gaussian regulator 
\cite{Loewe2}. This fact softens the
instabilities generated by the truncation of the pole series at
low temperature \cite{Buballa2}.\\

In this article we will develop the real time formalism for a nNJL model with a fractional Lorentzian regulator, which produces a cut in the complex plane, exhibiting a propagator with only complex poles. We will get the behavior of the masses of the quasiparticles as a function of the vacuum expectation value of a scalar bosonic field and we will look for a critical temperature at which chiral symmetry is restored in the chiral limit. We will then include a finite baryon chemical potential and look at how this affects the chiral symmetry restoration.\\

The paper is organized as follows. In Sec. II, we introduce the nNJL model and develop the real time formalism in a general manner. In Sec. III we will turn to 
our particular choice of regulator and study the behaviour of the masses as $T$ increases. We will also find the critical temperature for chiral symmetry restoration. In Sec. IV we present a brief discussion on the thermodynamical potential of the model and the incorporation of nonzero chemical potential. In Sec. V we present our conclusions.\\

\section{\lowercase{n}NJL Model in real time formalism.}

We consider the nNJL model, described by the Euclidean Lagrangian
\begin{multline}\mathcal{L}_E=\left[\bar{\psi}(x)(-i\slashed{\partial}+m)\psi(x)-\frac{G}{2}j_a(x)j_a(x)\right].\end{multline}
The nonlocal aspects of the model are introduced through the nonlocal currents $j_a(x)$ defined as
\begin{equation}j_a(x)=\int d^4y\,d^4z\,r(y-x)r(z-x)\bar{\psi}(x)\Gamma_a\psi(z),\end{equation} 
where $\Gamma_a=(1,i\gamma^5\vec{\tau})$. A standard bosonization procedure can be performed on the model by incorporating scalar ($\sigma$) and pseudoscalar ($\vec{\pi}$) fields. Quark fields can then be integrated out \cite{Scoccola5, Scoccola6}. We proceed in the mean field approximation. We assume the $\vec{\pi}$ fields to have null mean value because of isospin symmetry. So we take
\begin{eqnarray}
\sigma&=&\bar{\sigma}+\delta\sigma\\
\vec{\pi}&=&\delta\vec{\pi},
\end{eqnarray}
where $\bar{\sigma}$ is the vaccum expectation value of the scalar field. To first order in the fluctuations we can write the mean field effective action
\begin{equation}\Gamma^{MF}=V_4\left[\frac{\bar{\sigma}^2}{2G}-2N_c\int\frac{d^4q_E}{(2\pi)^4}\tr\ln S_E^{-1}(q_E)\right],\end{equation}
with $S_E(q_E)$ being the Euclidean effective propagator
\begin{equation}S_E=\frac{-\slashed{q}_E+\Sigma(q_E^2)}{q_E^2+\Sigma^2(q_E^2)}.\end{equation}
Here, $\Sigma(q_E^2)$ is the constituent quark mass
\begin{equation}\Sigma(q_E^2)=m+\bar{\sigma}r^2(q_E^2).\end{equation}
Finite temperature ($T$) and chemical potential ($\mu$) effects can be incorporated through the Matsubara formalism. To do so, we make the following substitutions
\begin{eqnarray}
V_4&\rightarrow&V/T\\
q_4&\rightarrow&-q_n\\
\int\frac{dq_4}{2\pi}&\rightarrow&T\sum_n,
\end{eqnarray}
where $q_n$ includes the Matsubara frequencies and the chemical potential
\begin{equation}q_n\equiv(2n+1)\pi T+i\mu.\end{equation}
 
The grand canonical thermodynamical potential in the mean field approximation is given by $\Omega_{MF}(\bar{\sigma},T,\mu)=(T/V)\Gamma_{MF}(\bar{\sigma},T,\mu)$ \cite{Kapusta}. The value of $\bar{\sigma}$ can then be obtained through the solutions of the gap equation $\partial\Omega_{MF}/\partial\bar{\sigma}=0$, which means
\begin{equation}
\left.\frac{\bar{\sigma}}{G}=2N_cT\sum_n\int\frac{d^3q}{(2\pi)^3}
r^2(q_E^2)\tr 
S_E(q_E)
\right|_{q_4=-q_n}.
\label{gap}
\end{equation}
So far we have worked in the imaginary time formalism. In order to go to the real time formalism we must perform a rotation from Euclidean to Minkowski space by taking $q_4=iq_0$. We will then obtain the zero temperature propagator in Minkowski space
\begin{equation}S_0=i\frac{\slashed{q}+\Sigma(-q^2)}{q^2-\Sigma^2(-q^2)},\label{zp}\end{equation}
where $q^2=-q_E^2$. This propagator has singularities in the complex plane $q^2$. In what follows we will assume that this propagator has only complex singularities. Following our quasiparticle intepretation of the singularities, we will define a mass and a decay width by writing the poles of the propagator at
\begin{equation}q^2=\mathcal{M}^2=M^2+iM\Gamma,\end{equation} 
where $M$ is the constituent mass of the quasiparticle and $\Gamma$ its decay width. Our next step is to introduce the thermal propagator in the real time formalism.\\

In the real time formalism, the number of degrees of freedom is doubled \cite{Ojima1, Ojima2, Kobes, Landsman, LeBellac, Das}. This means that the thermal propagator is given by a $2\times2$ matrix with elements $S_{ij}$. However, in one-loop calculations only the $S_{11}$ component is necesary. We can write a general expression for $S_{11}$ in terms of the spectral density function (SDF)
\begin{equation}S_{11}=\int\frac{dk_0}{2\pi i}\frac{\rho(k_0,\boldsymbol{q})}{k_0-q_0-i\varepsilon}-n_F(q_0-\mu)\rho(q),\label{prop}\end{equation}
where $n_F(q_0-\mu)$ is the Fermi-Dirac distribution $n_F(q_0-\mu)=({\rm{e}}^{(q_0-\mu)/T}+1)^{-1}$. We can obtain the SDF from
\begin{equation}\rho(q)=S_+(q)-S_-(q),\end{equation}
where
\begin{equation}S_{\pm}(q)=\pm\oint_{\Gamma^{\pm}}\frac{dz}{2\pi i}\frac{S_0(z\mp i\varepsilon,\boldsymbol{q})}{z-q_0\pm i\varepsilon}.\end{equation}
\vspace{0.2cm}\\
This is just a generalization of the free particle case where $\rho(q)=S_0(q_0+i\varepsilon,\boldsymbol{q})-S_0(q_0-i\varepsilon,\boldsymbol{q})$. The integration path $\Gamma^{\pm}$ is shown in Fig. 1.\\

\begin{figure}[!htb]
\begin{center}
\includegraphics[scale=0.4]{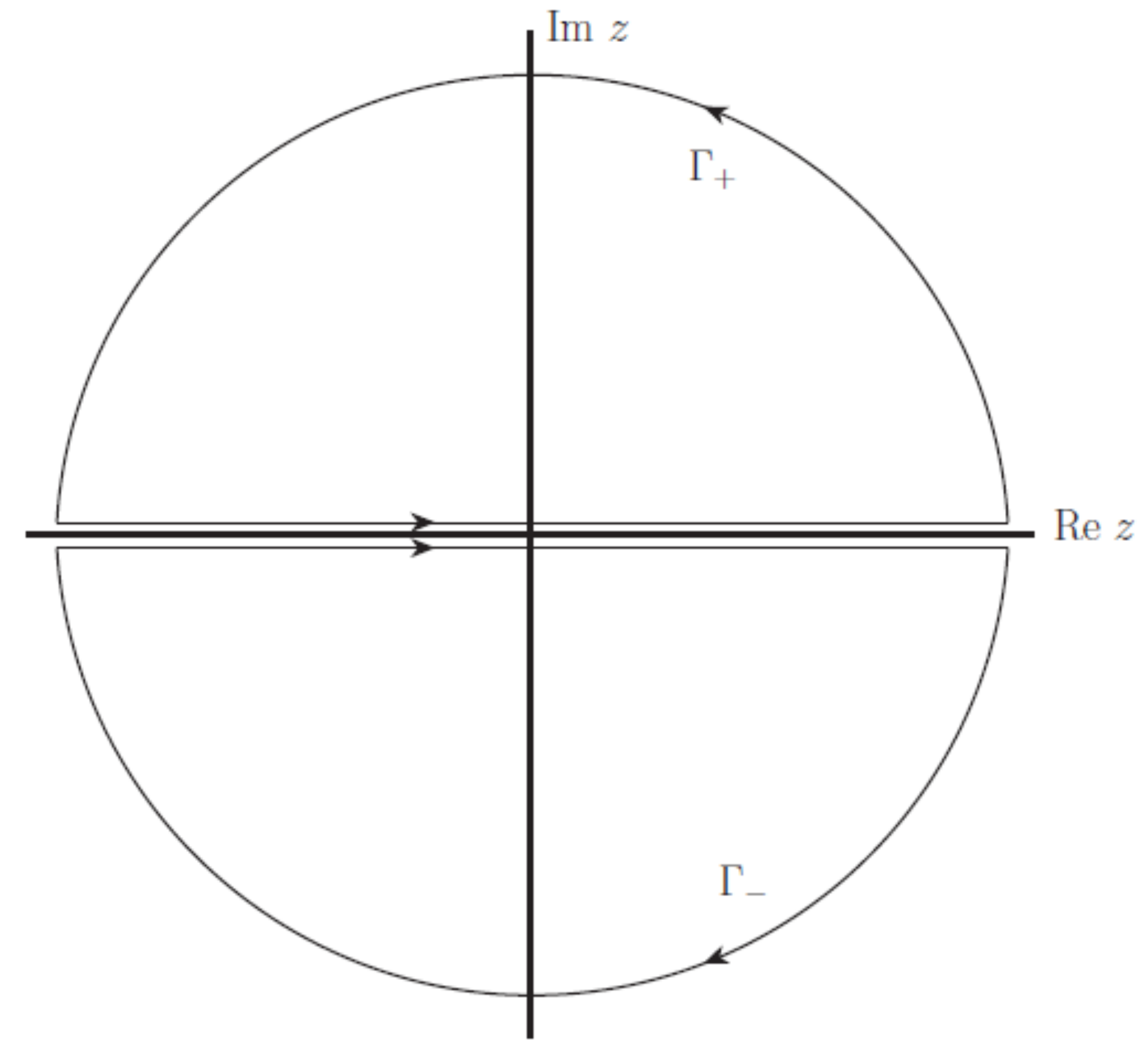}
\caption{Integration path in the definition of $S_{\pm}$.}
\label{pathS}
\end{center}
\end{figure}

Performing the integrations we get

\begin{multline}\rho(q)=\sum_{\mathcal{M}}\frac{i}{(\mathcal{M}^2-q^2)((\mathcal{M}^2)^*-q^2)}\\\times\left[((\mathcal{M}^2)^*-q^2)A(\mathcal{M}^2)-(\mathcal{M}^2-q^2)A((\mathcal{M}^2)^*)\right]\end{multline}
where the sum is over the various poles ($\mathcal{M}$) of the propagator and
\begin{multline}A(\mathcal{M}^2)=\frac{Z(\mathcal{M}^2)}{2E}\left(q_0(\slashed{q}+\Sigma(-\mathcal{M}^2))\right.\\\left.-\gamma^0(q^2-\mathcal{M}^2)\right).\end{multline}
As usual, $E^2=\mathcal{M}^2+\boldsymbol{q}^2$ and
\begin{equation}Z(\mathcal{M}^2)=\left.\left[\frac{\partial}{\partial q^2}\left(q^2-\Sigma^2(-q^2)\right)\right]^{-1}\right|_{q^2=\mathcal{M}^2},\label{Z}\end{equation}
is the renormalization constant. The finite temperature propagator can then be obtained by putting this result into Eq. (\ref{prop}). Finite temperature and chemical potential contributions to this propagator will be decoupled from the zero temperature ones. In this sense, we can write our propagator as
\begin{equation}S_{11}(q,T,\mu)=S_0(q)+\tilde{S}(q,T,\mu).\end{equation}

Here all finite temperature effects are inside $\tilde{S}(q,T,\mu)$ and $S_0(q)$ is just the zero temperature propagator. From Eq. (\ref{prop}) we can see that all of the $T,\mu$ contribution comes from the second term. However, this term does not vanish when $T,\mu\rightarrow0$ since $n_F(q_0-\mu)\rightarrow\theta(-q_0)$. To avoid this, we can define
\begin{equation}\tilde{S}(q,T,\mu)=-n_F(q_0-\mu)\rho(q)+K(q),\end{equation}
where we have added the function $K(q)$ that is fixed in order to ensure that $\tilde{S}(q,0,0)=0$. Our next step is to obtain the gap equation in real time formalism. We can achieve this by taking $S_E\rightarrow S_{11}$ in Eq. (\ref{gap}). In this manner we get
\begin{equation}\frac{\partial\Omega_{MF}}{\partial\bar{\sigma}}=g_0(\bar{\sigma})+\tilde{g}(\bar{\sigma},T,\mu)=0,\label{potential}\end{equation}
where
\begin{eqnarray}
g_0(\bar{\sigma})=\frac{\bar{\sigma}}{G}-\frac{N_c}{\pi^2}\int_0^\infty dq_Eq_E^3\frac{r^2(q_E^2)\Sigma(q_E^2)}{q_E^2+\Sigma^2(q_E^2)}\\
\tilde{g}(\bar{\sigma},T,\mu)=-2N_c\int\frac{d^4q}{(2\pi)^4}r^2(-q^2)\tr\tilde{S}(q,T,\mu)\label{gtilde},
\end{eqnarray}
and where again $\tilde{g}(\bar{\sigma},T,\mu)$ has all of the finite temperature and chemical potential contributions to the gap equation. By putting our expression for $\tilde{S}$ into Eq. (\ref{gtilde}) we get
\begin{multline}\tilde{g}(\bar{\sigma},T,\mu)=2iN_c\sum_{\mathcal{M}}Z(\mathcal{M}^2)\Sigma(-\mathcal{M}^2)\int\frac{d^4q}{(2\pi)^4}\frac{r^2(-q^2)}{E}\\\times\left(n_F(q_0-\mu)+n_F(q_0+\mu)\right)\\\times\left[\frac{q_0}{\mathcal{M}^2-q^2}-\frac{q_0}{(\mathcal{M}^2)^*-q^2}\right].\end{multline}
The integration in $q_0$ can be performed along the path shown in Fig. 2.

\begin{figure}[!htb]
\begin{center}
\includegraphics[scale=0.4]{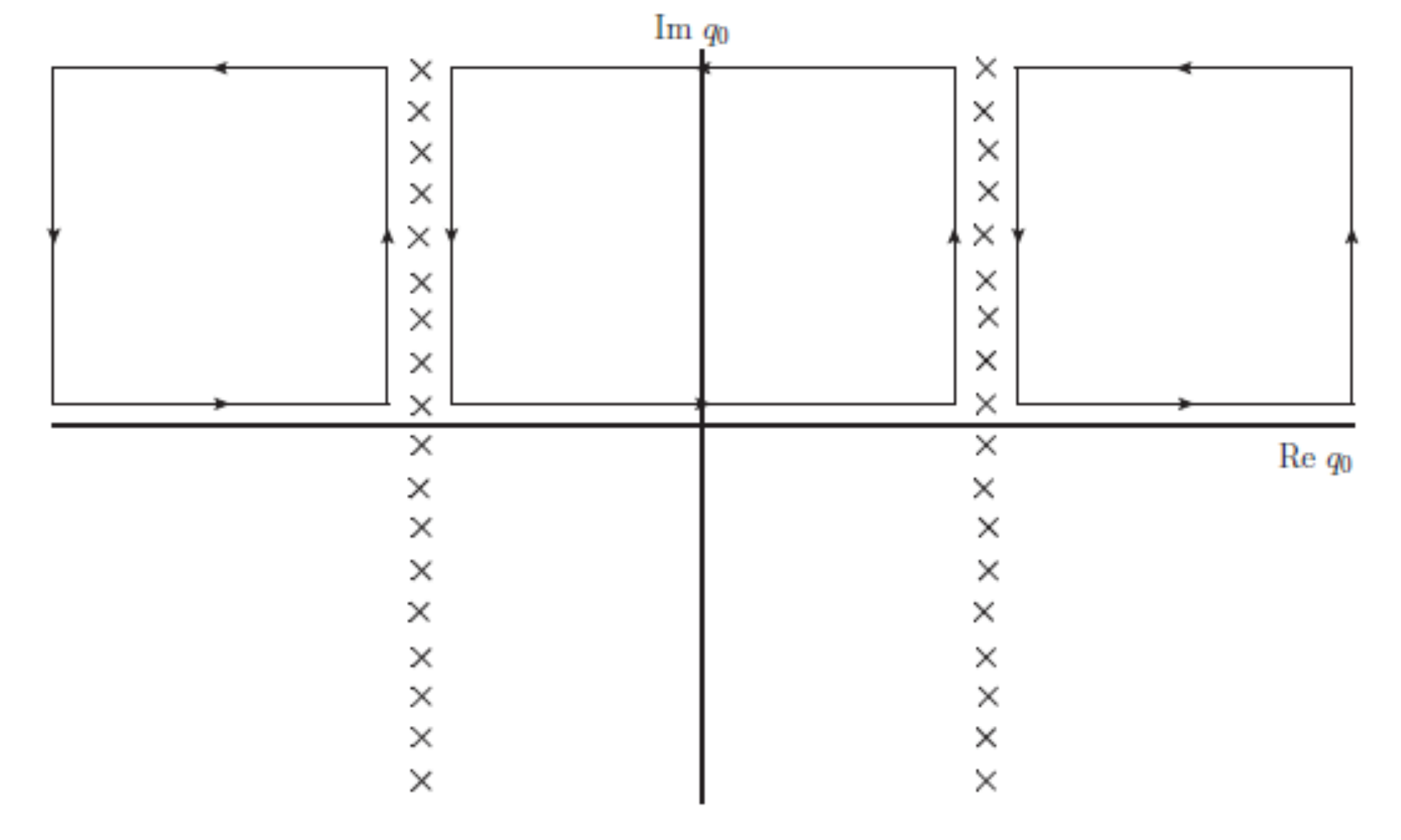}
\caption{Integration path for the thermal part of the gap equation. The poles of the Fermi-Dirac distribution are marked with crosses}
\end{center}
\end{figure}

The integration can be computed to give
\begin{multline}\tilde{g}(\bar{\sigma},T,\mu)=-\frac{N_c}{\pi^2}\sum_{\mathcal{M}}\left[Z(\mathcal{M}^2)\Sigma(-\mathcal{M}^2)r^2(-\mathcal{M}^2)\right.\\\times\int dkk^2\frac{n_F(E-\mu)+n_F(E+\mu)}{E}\\+\left.\left(\mathcal{M}^2\rightarrow(\mathcal{M}^2)^*\right)\right].\label{gapfin}\end{multline}
With this we have temperature and chemical potential dependent expressions for our propagator and the gap equation. In the next section we will choose a regulator and look for the critical temperature at which chiral symmetry is restored.

\section{Poles of the propagator and chiral symmetry restoration.}

So far we have not said much about the regulator $r(q_E^2)$. In \cite{Scoccola1} a regulator has been constructed that agrees quite well with lattice data. Inspired by this fact, we choose our regulator to be
\begin{equation}r^2(q_E^2)=\frac{1}{1+\left(\frac{q_E^2}{\Lambda^2}\right)^{3/2}},\end{equation}
where $\Lambda$ is a free parameter of the model to be determined. Also, we will work in the chiral limit where $m=0$. We can put this regulator in Eq. (\ref{Z}) to get 
\begin{equation}Z(\mathcal{M}^2)=\left[1-3\frac{\left(-\mathcal{M}^2/\Lambda^2\right)^{3/2}}{1+\left(-\mathcal{M}^2/\Lambda^2\right)^{3/2}}\right]^{-1}.\end{equation}
To complete the description of our model, we need to fix its free parameters. Since we are working in the chiral limit, we are left with only three parameters that need to be fixed, namely $G, \Lambda$ and $\bar{\sigma}$ at $T=0$.\\

The value of $\bar{\sigma}$ at $T=0$ can be determined from the gap equation in eq. (\ref{gap}). In order to fix the other two parameters we resort to quantities of known value: the chiral condensate at zero temperature and the pion decay constant. It is quite easy to obtain an expression for the chiral condensate at zero temperature
\begin{equation}
\langle q\bar{q}\rangle=-N_c\int\frac{d^4q_E}{(2\pi)^4}\tr 
S_E(q_E) -\{\bar\sigma\to 0\}.
\end{equation}
In our notation, $\langle q\bar{q}\rangle$ includes only one flavor, i.e. $\langle q\bar{q}\rangle=\langle u\bar{u}\rangle=\langle d\bar{d}\rangle$. Finally, we need an expression for the pion decay constant in the chiral limit. Such an expression can be obtained from the quadratic terms in the mean field expansion of the action \cite{Scoccola2}
\begin{equation}f_\pi^2=2N_c\int\frac{d^4q_E}{(2\pi)^4}\frac{2\Sigma^2(q_E)-q_E^2\Sigma(q_E)\Sigma^\prime(q_E)}{[q_E^2+\Sigma^2(q_E^2)]}.\end{equation}
We take $(-\langle q\bar{q}\rangle)^{1/3}=260$ MeV and $f_\pi=90$ MeV. With this input, we obtain for our parameters
\begin{eqnarray}
\Lambda&\approx&635\mbox{ MeV}\\
\bar{\sigma}&\approx&261\mbox{ MeV}\\
G&\approx&28\cdot10^{-6}\mbox{ MeV}{}^{-2}.
\end{eqnarray}
Once we have fixed the parameters, we can work with our model and study its properties in the real time formalism.\\

In Minkowski space our regulator will take the form
\begin{equation}r^2(-q^2)=\frac{1}{1+\left(-\frac{q^2}{\Lambda^2}\right)^{3/2}}.\label{reg}\end{equation}
We should now define what we will understand by the semi-integer exponent in the previous equation. This regulator was originally defined in Euclidean space, in which case, the function $z^{3/2}$ is defined within the real numbers and is well behaved. However, once we have rotated to real time, we need to define this function in the complex plane. In this case, the function has a cut in the complex plane being is a multivalued function. Usual definitions of such a function are made in such a way that it will no longer be a multivalued function, however, this is a feature we want to keep, so we will define the function as
\begin{equation}z^{3/2}=\left(R{\rm{e}}^{i\theta}\right)^{3/2}=r^{3/2}{\rm{e}}^{\frac{3}{2}i\theta}.\end{equation}
This means that we will keep the multivalued nature of our regulator, which will double the number of singularities our propagator will have (for each singularity in the first Riemmann sheet we will get another one in the second sheet). We can search for such singularities (poles) in our propagator by considering the solutions to
\begin{equation}q^2-\Sigma^2(-q^2)=0.\label{polos}\end{equation}

In this manner we find eight poles which, however, appear in complex conjugates pairs so we can speak of only four poles plus their complex conjugates. We then have four different masses and decay widths at
\begin{equation}\mathcal{M}_j^2=M_j^2\pm iM_j\Gamma_j,\end{equation}
with $j=1,\ldots,4$. 

\begin{figure}[!htb]
\begin{center}
\includegraphics[scale=0.2]{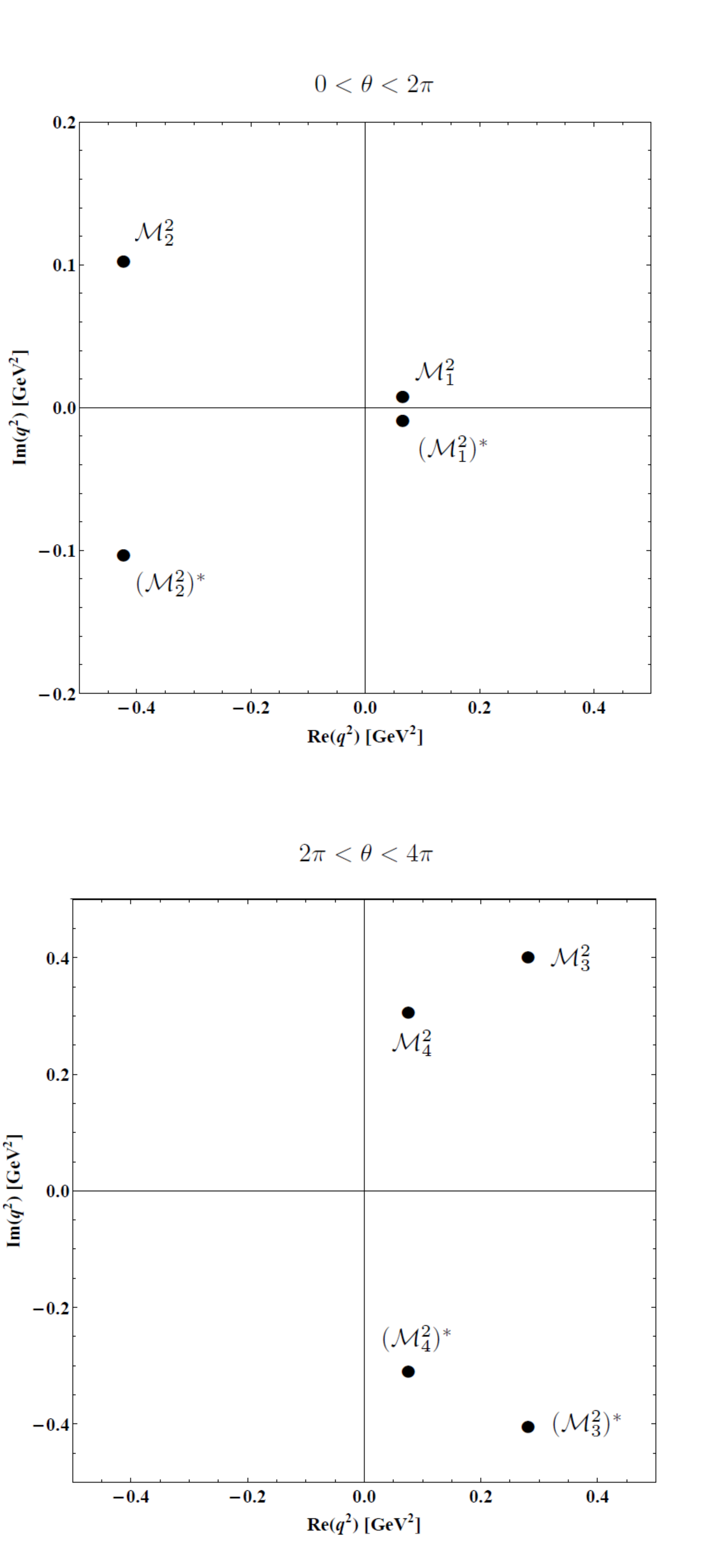}
\end{center}
\caption{Position of the poles in both Riemann sheets for $T=0$ ($\bar{\sigma}=261$ MeV). The dots mark where the poles are.}
\label{Hoja}
\end{figure}

We can parametrize the poles using their complex argument as $q^2=R(\theta){\rm{e}}^{i\theta}$. We can get an expression for  $R(\theta)$ from Eq. (\ref{polos}). In this way, our poles are completely described by their argument $\theta$. Our propagator then has two Riemann sheets, one for $0<\theta<2\pi$ and another for $2\pi<\theta<4\pi$. With this notation and using Eq. (\ref{polos}) we can plot where these poles are in both sheets.\\

As can be seen from Fig. \ref{Hoja} all of the singularities have nonvanishing imaginary parts. However, the pole at $q^2=\mathcal{M}_2^2$ has $M_2^2<0$. Following our interpretation this would be a particle with a negative square mass. Such a particle is a highly unstable one and it could not contribute to a condensate.  Including such a quasiparticle would lead to nonphysical results, like a condensate that grows with $T$. Therefore we will not consider it in the further analysis.\\

We can also use Eq. (\ref{polos}) to get the behaviour of the poles as a function of $\bar{\sigma}$.

\begin{figure}[!h]
\begin{center}
\includegraphics[scale=0.2]{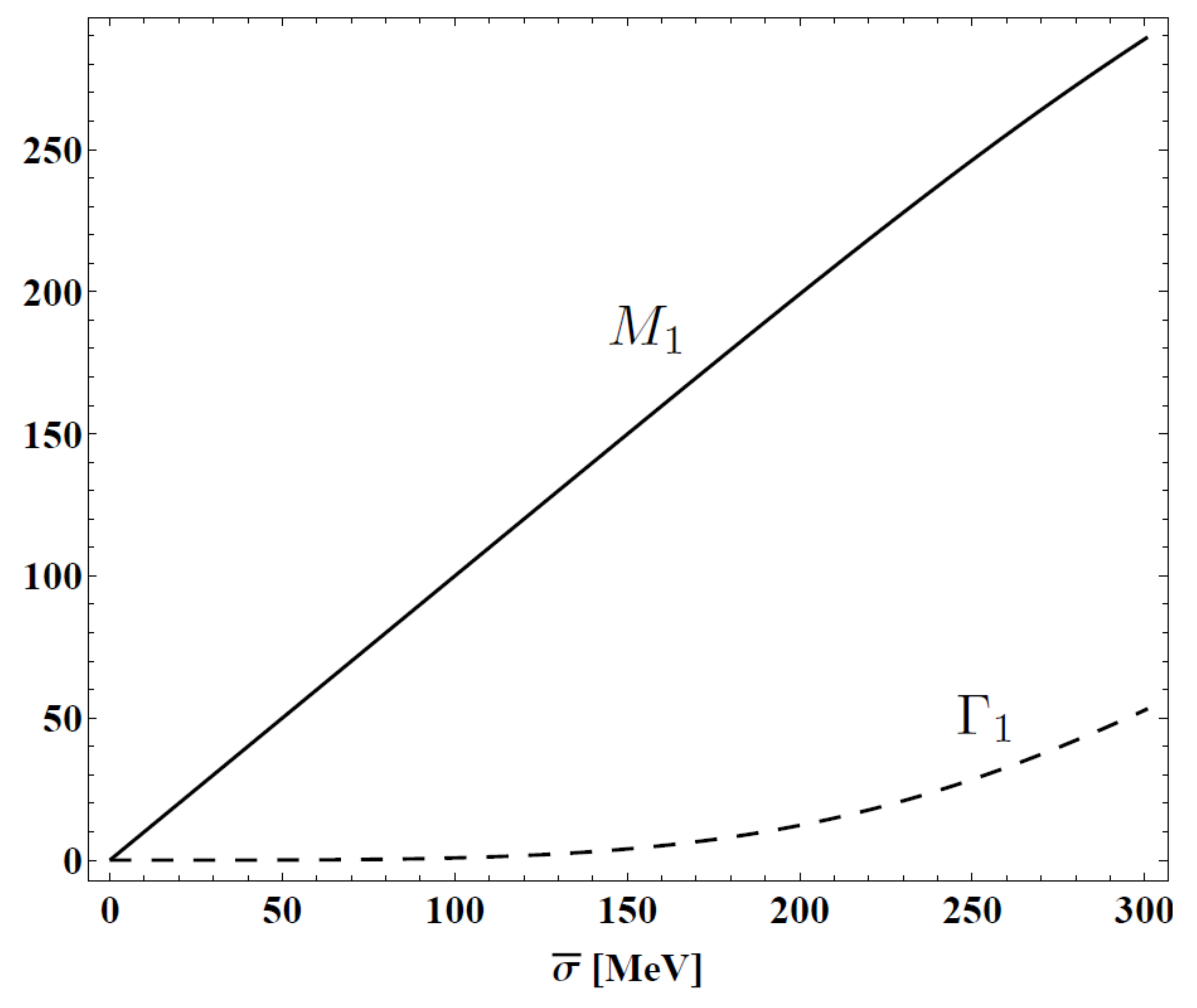}
\end{center}
\caption{Behaviour of the pole at $q^2=\mathcal{M}_1^2$ as a function of $\bar{\sigma}$. The solid line stands for $M_1$ and the dashed line for $\Gamma_1$. All quantities are given in MeV.}
\label{Polo11}
\end{figure}

\begin{figure}[!h]
\begin{center}
\includegraphics[scale=0.2]{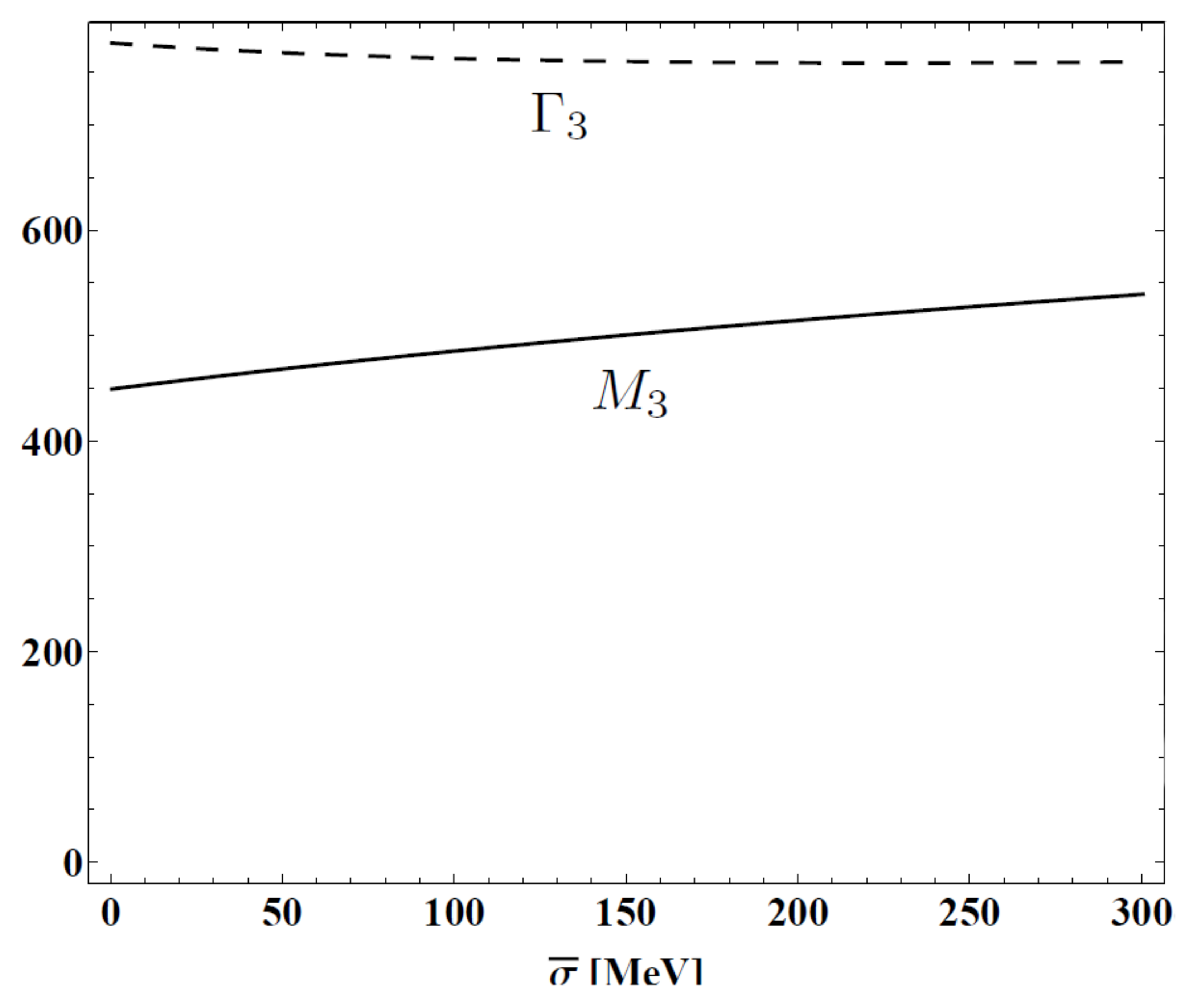}
\end{center}
\caption{Behaviour of the pole at $q^2=\mathcal{M}_3^2$ as a function of $\bar{\sigma}$. The solid line stands for $M_3$ and the dashed line for $\Gamma_3$. All quantities are given in MeV.}
\label{Polo22}
\end{figure}

\begin{figure}[!h]
\begin{center}
\includegraphics[scale=0.2]{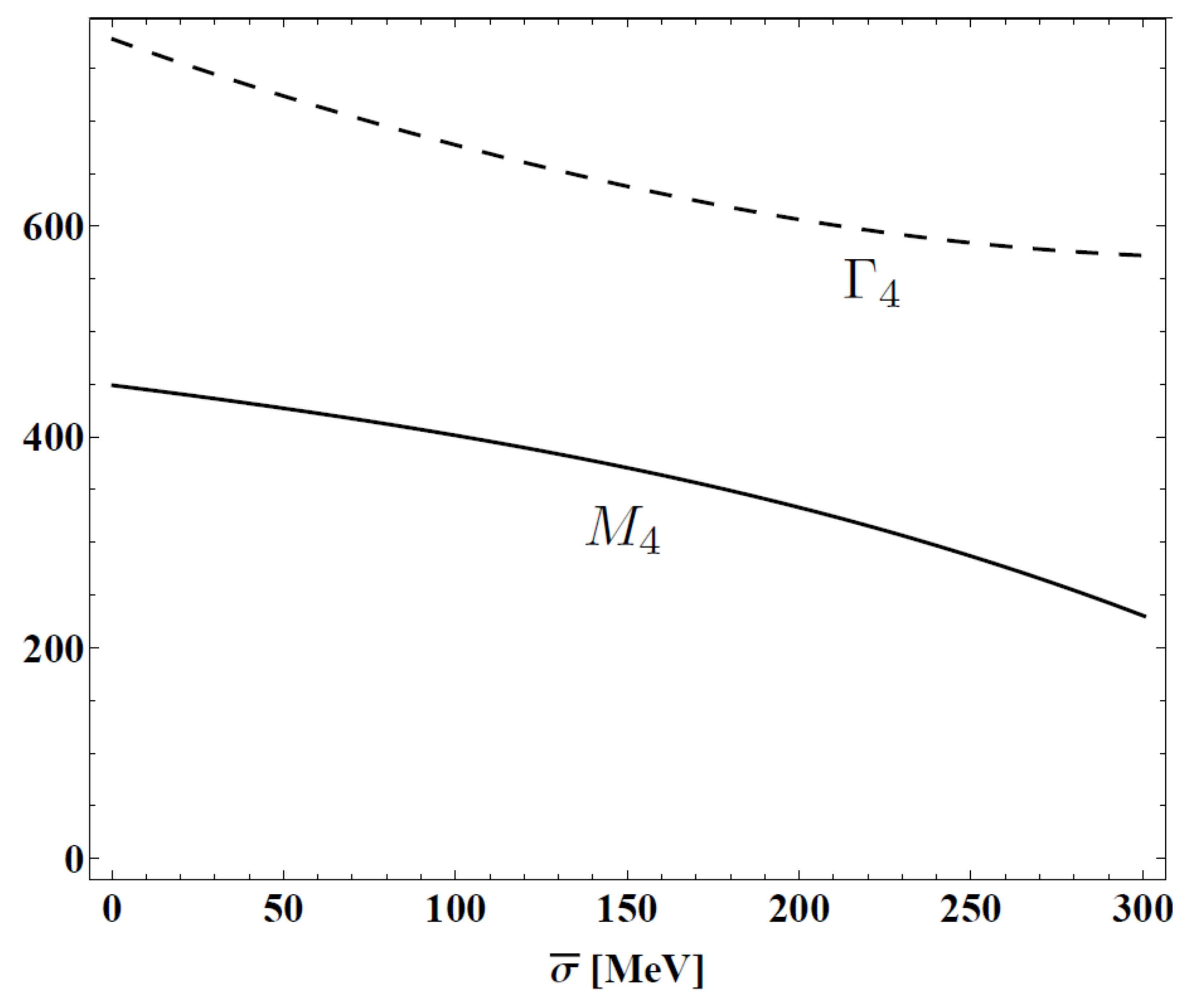}
\end{center}
\caption{Behaviour of the pole at $q^2=\mathcal{M}_4^2$ as a function of $\bar{\sigma}$. The solid line stands for $M_4$ and the dashed line for $\Gamma_4$. All quantities are given in MeV.}
\label{Polo44}
\end{figure}

As can be seen from Figs. \ref{Polo11}, \ref{Polo22} and \ref{Polo44}, for low temperatures (high $\bar{\sigma}$) the three remaining poles have similar masses lying between 200 and 500 MeV. However for high temperatures (near the critical temperature for chiral symmetry restoration and low $\bar{\sigma}$) the pole at $q^2=\mathcal{M}_1^2$ (Fig. \ref{Polo11}) has $M_1\rightarrow0$ while the other two remain with $M_{3,4}>400$ MeV. These other two singularities also have a much greater decay width. Here, the quasiparticle interpretation we have in the real time formalism comes in handy. We can interpret these two singularities as a much more massive and short-lived quasiparticles with respect to that of Fig. \ref{Polo11}. Such quasiparticles should not make a significant contribution to a condensate. Because of these two reasons  their contribution to the condensate is neglegible. \\

We can also use eq. (\ref{gapfin}) to get the behaviour of $\bar{\sigma}$ as a function of temperature for $\mu=0$ and with this we can plot the behaviour of the mass and decay width of the remaining quasiparticle as a function of $T$.

\begin{figure}[!htb]
\begin{center}
\includegraphics[scale=0.2]{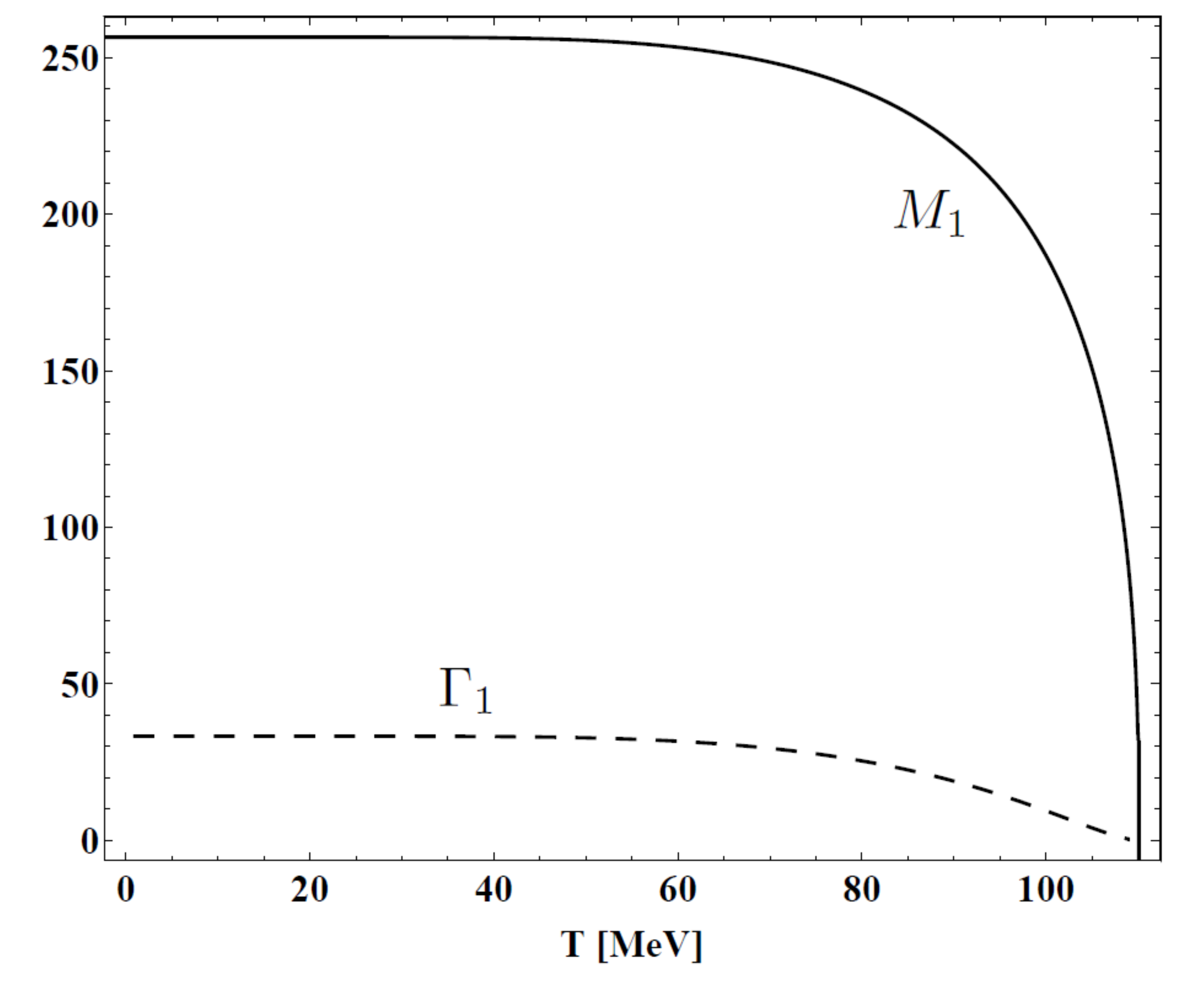}
\end{center}
\caption{Behaviour of the pole as a function of temperature. The solid line stands for $M_1$ and the dashed line stands for $\Gamma_1$. All quantities are given in GeV.}
\label{Polo1}
\end{figure}

As we can see from Fig. \ref{Polo1}, for high enough temperature the mass of the pole rapidly decreases. Because of this small mass, this pole has a significant contribution to the condensate, while the $\mathcal{M}_3$ and $\mathcal{M}_4$ ones are neglegible.\\

We can use the solutions of the gap equation to compute the chiral condensate
\begin{equation}
\langle q\bar{q}\rangle=-N_c\int\frac{d^4q}{(2\pi)^4}\tr 
S_{11}(q,T,\mu) -\{\bar\sigma\to 0\}.
\end{equation}
This can be easily computed and we can obtain the critical temperature at which $\langle q\bar{q}\rangle=0$, i.e. the temperature at which chiral symmetry is restored.

\begin{figure}[!htb]
\begin{center}
\includegraphics[scale=0.2]{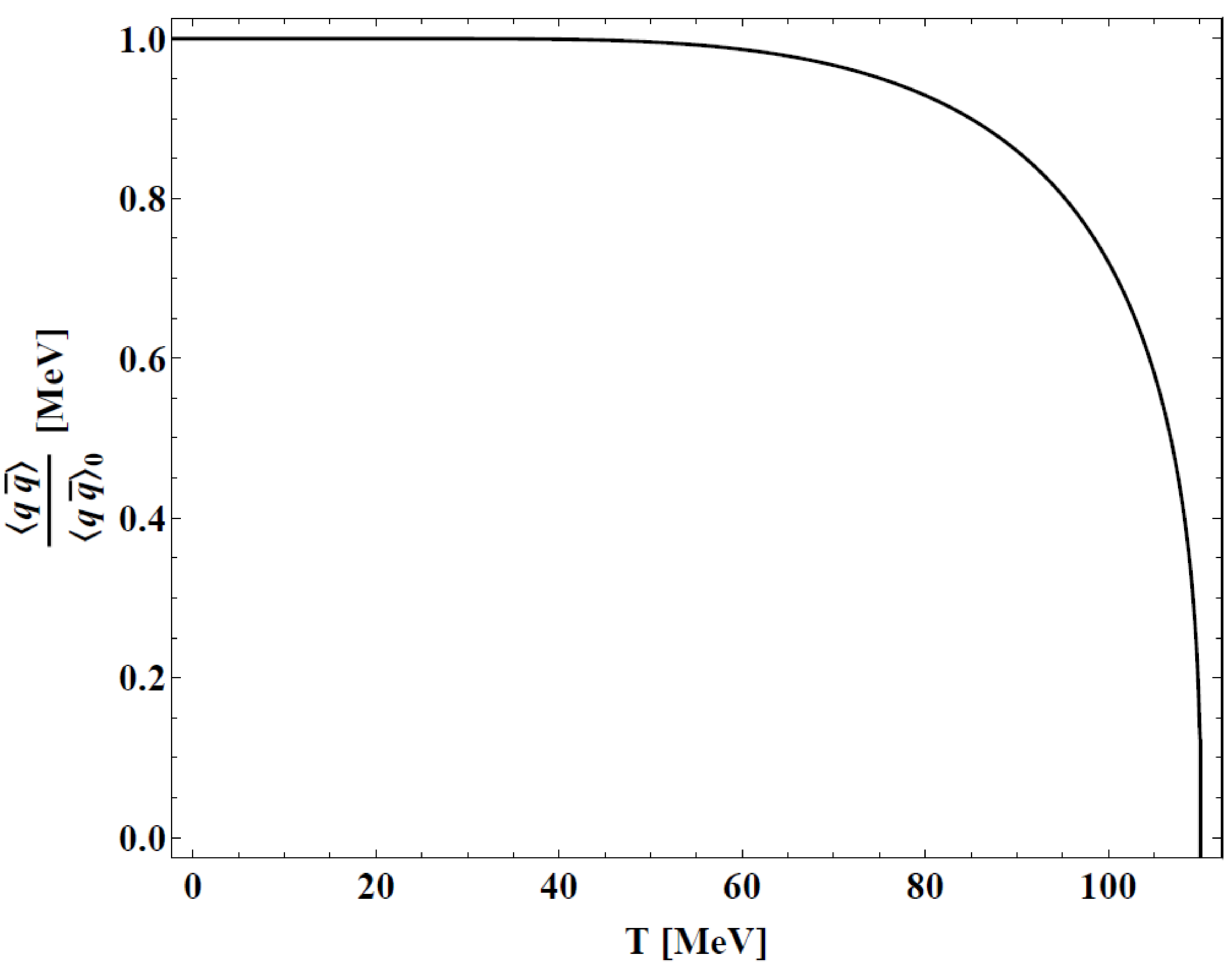}
\end{center}
\caption{Behaviour of $\langle q\bar{q}\rangle$ as a function of temperature. $\langle q\bar{q}\rangle_0$ stands for the chiral condensate at zero temperature.}
\label{chiral}
\end{figure}

As can be seen from Fig. \ref{chiral} a critical temperature is found around $T\approx110$ MeV.  This is a reasonable result since similar temperatures are found in models beyond the chiral limit and with Polyakov loop included \cite{Scoccola3, Scoccola4}. The critical temperature is not the same . It is important to note that such a transition would not be found if we had included the singularity with $M_2^2<0$. Since it has a negative real part for the square mass, it is a highly unstable particle and hence, it cannot contribute to a condensate. If we had not dropped the negative square mass singularity we would not have found a chiral symmetry restoration. This is the reason why it is important to analize the behaviour of the poles of the propagator and the squared masses that come from it.\\

The extension of this model beyond the chiral limit implies the existence of 
more poles. The inclusion of the Polyakov loop can be done easily using the 
Polyakov gauge  \cite{Scoccola20, Ratti}. In this scenario we would find even 
more poles due to the nature of the inverse propagator matrix.  

\begin{figure}[!htb]
\begin{center}
\includegraphics[scale=0.22]{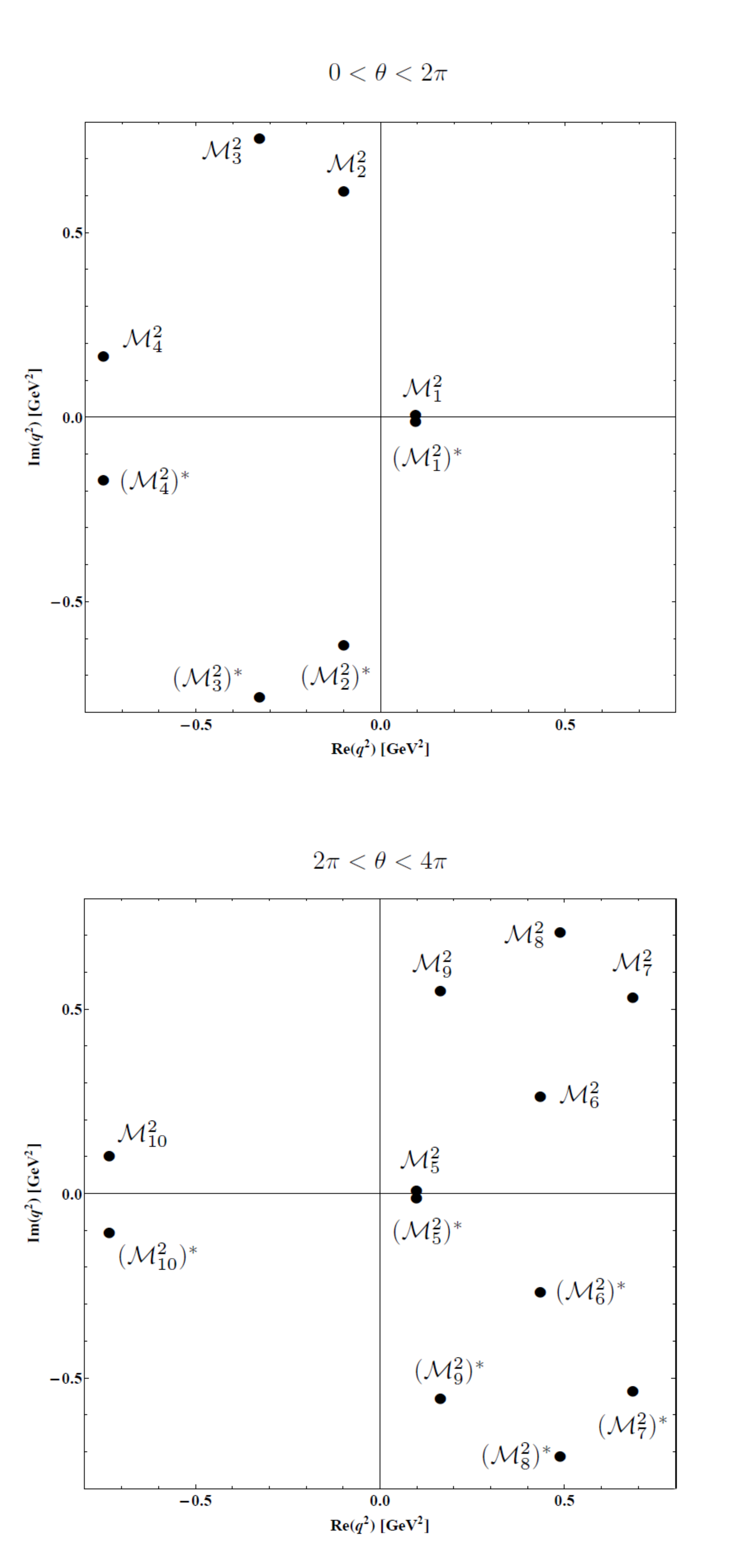}
\end{center}
\caption{Position of the poles in both Riemann sheets for $T=0$ beyond the 
chiral limit. The dots mark the location of the poles.}
\label{Polosnew}
\end{figure}

Figure \ref{Polosnew} shows the position of the poles of the propagator for 
$T=0$ beyond the chiral limit. The appearance of more poles is fairly easy to 
take into account. The formalism can be worked out in exactly the same way. We 
only would have more terms in each expression, but they would all have the same 
form, just being evaluated at different poles. 
The inclusion of Polyakov loops, however, could be irrelevant since the model 
includes confinement effects by itself. Nevertheless, the inclusion of the Polyakov loop
contributes to eliminate some instabilities that appear in
regulators that exhibit complex poles \cite{Buballa2, Blaschke3}.

\section{Thermodynamical potential and chemical potential}

We will now compute the grand canonical thermodynamical potential. We start from Eq. (\ref{potential}) and integrate to get
\begin{equation}\Omega_{MF}=\Omega_0(\bar{\sigma})+\tilde{\Omega}(\bar{\sigma},T,\mu),\end{equation}
where
\begin{eqnarray}
\Omega_0(\bar{\sigma})&=&\int g_0(\bar{\sigma})d\bar{\sigma}\\
\tilde{\Omega}(\bar{\sigma},T,\mu)&=&\int\tilde{g}(\bar{\sigma},T,\mu)d\bar{\sigma}+C(T,\mu),\label{tilde}
\end{eqnarray}
and $C(T,\mu)$ is an integration constant that we will choose in order to satisfy $\Omega_{MF}(\bar{\sigma}=0)=0$. It is a straightforward excercise to get
\begin{equation}\Omega_0=\frac{\bar{\sigma}^2}{2G}-\frac{N_c}{2\pi^2}\int_0^\infty dq_E q_E^3\ln\left[q_E^2+\Sigma^2(q_E^2)\right].\end{equation}
The computation of $\tilde{\Omega}$ is less trivial. We want to integrate in $\bar{\sigma}$, however, $\tilde{g}(\bar{\sigma},T,\mu)$ is written in such a way that the $\bar{\sigma}$ dependence is hidden on the pole ($\mathcal{M}$) dependence. Since $\mathcal{M}$ is a pole of the propagator, we can write
\begin{equation}\Sigma^2(-\mathcal{M}^2)=\bar{\sigma}^2r^4(-\mathcal{M}^2)=\mathcal{M}^2.\end{equation}
Differentitating the previous equation, we find
\begin{equation}d\bar{\sigma}=\frac{d\mathcal{M}^2}{2r^2(-\mathcal{M}^2)\Sigma(-\mathcal{M}^2)Z(\mathcal{M}^2)}.\end{equation}
Putting this into Eq. (\ref{tilde}) we get
\begin{multline}\tilde{\Omega}(\bar{\sigma},T,\mu)=\frac{N_c}{2\pi^2}\sum_{\mathcal{M}}\int d\mathcal{M}^2\\\times\left[dkk^2\frac{n_F(E-\mu)+n_F(E+\mu)}{E}\right]\\+\left(\mathcal{M}^2\rightarrow(\mathcal{M}^2)^*\right)+C(T,\mu).\end{multline}
Finally, performing the $\mathcal{M}^2$ integration, we get
\begin{multline}\tilde{\Omega}=\frac{N_c}{\pi^2}\sum_{\mathcal{M}}\int dkk^2\left[2E-T\ln\left(1+{\rm{e}}^{\frac{E-\mu}{T}}\right)\right.\\\left.-T\ln\left(1+{\rm{e}}^{\frac{E+\mu}{T}}\right)\right]\\+\left(\mathcal{M}^2\rightarrow(\mathcal{M}^2)^*\right)+C(T,\mu).\end{multline}
From this expression we can compute the thermodynamical potential for different values of $\{T,\mu\}$.

\begin{figure}[!htb]
\begin{center}
\includegraphics[scale=0.22]{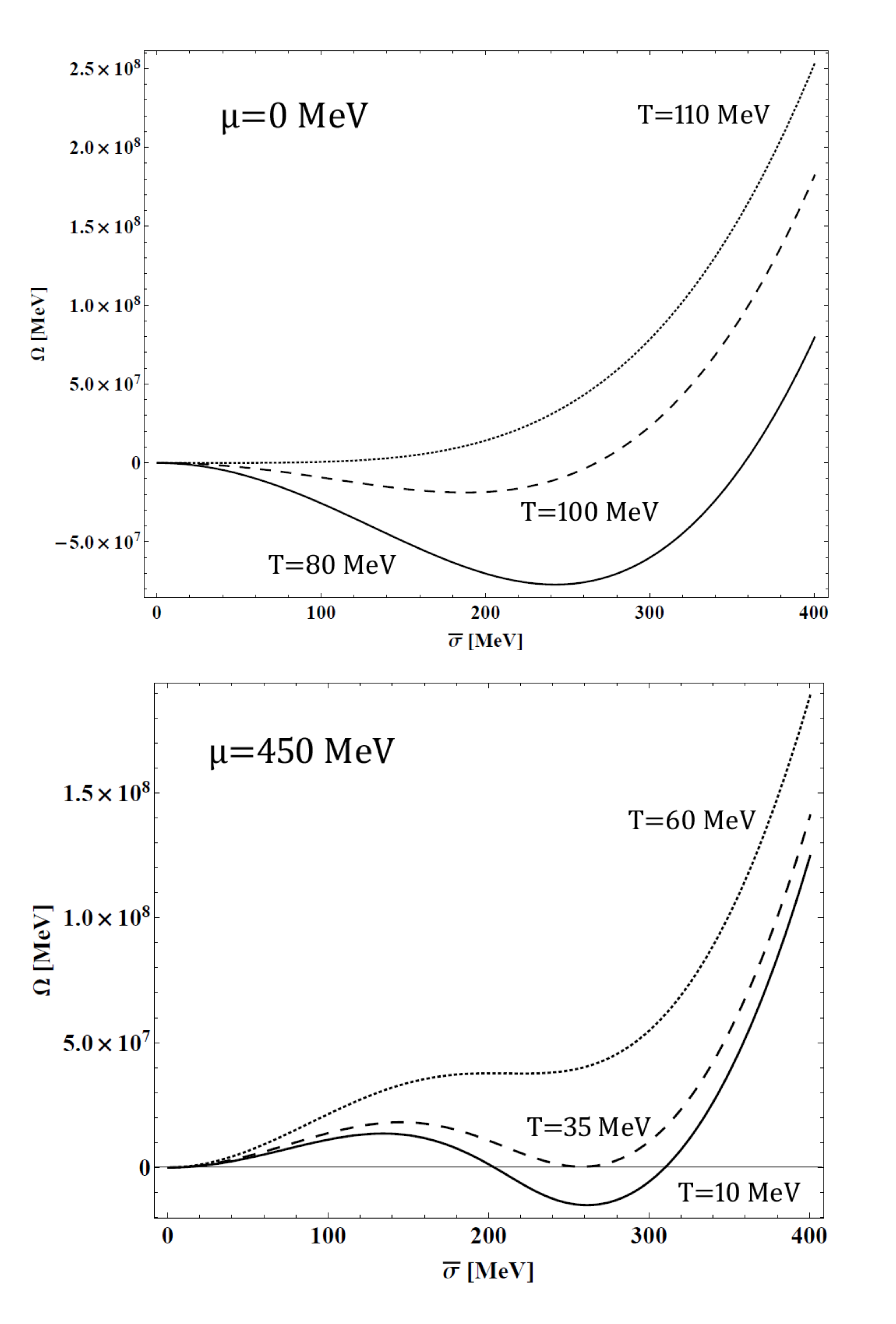}
\end{center}
\caption{Thermodynamical potential for $T$=(80, 100, 110 MeV) and $\mu=0$ MeV (top) and for $T$=(10, 35, 60 MeV) and $\mu=450$ MeV (bottom).}
\label{Chem5}
\end{figure}

As can be seen from the top plot in Fig. \ref{Chem5}, we have a second order phase transition around $T=110$ MeV for $\mu=0$. For higher values of the chemical potential, we find a first order phase transition and the critical temperature decreases. This computation can be extended to the whole $T-\mu$ plane ontaining then a phase diagram.

\begin{figure}[!htb]
\begin{center}
\includegraphics[scale=0.21]{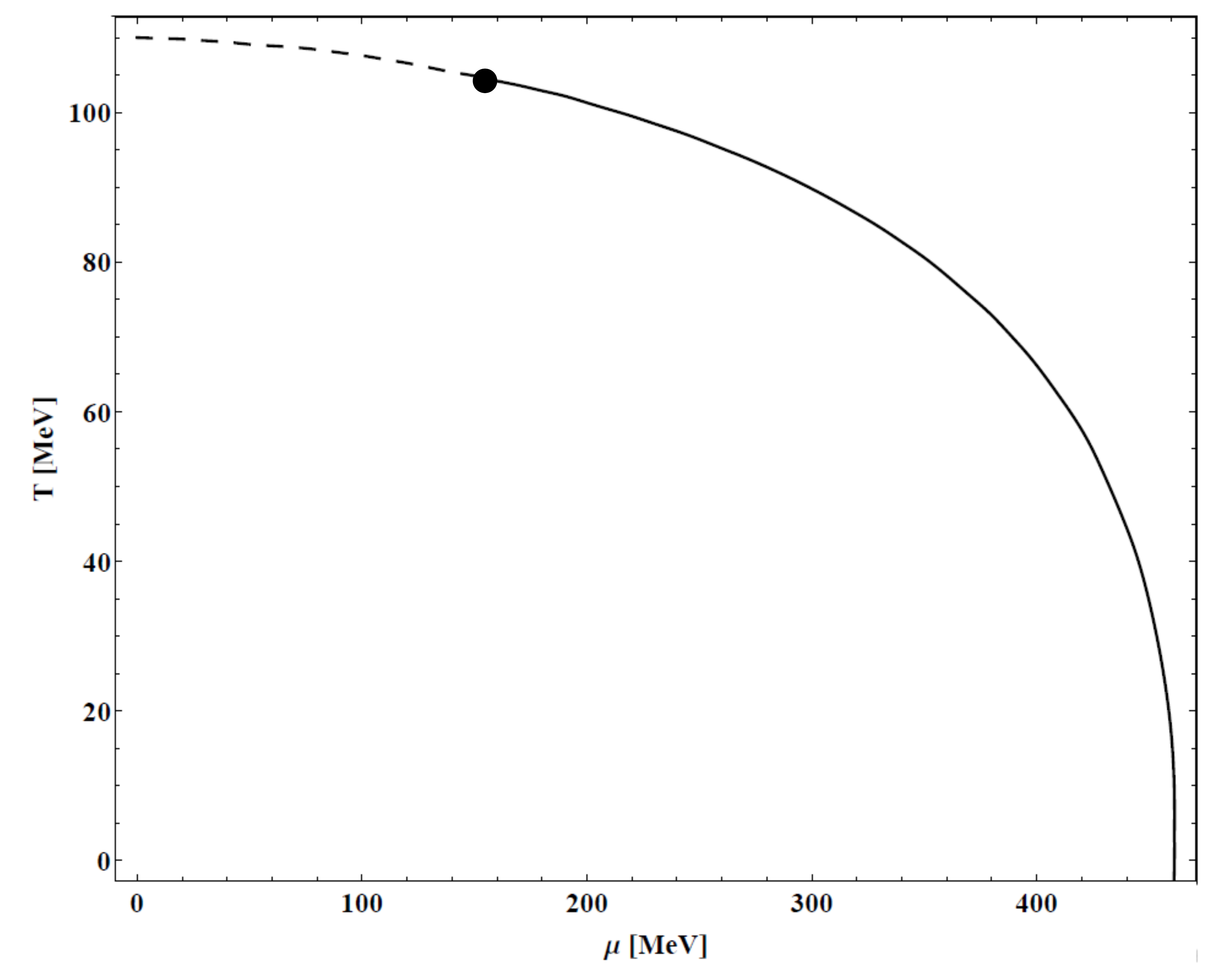}
\end{center}
\caption{$(T,\mu)$ phase diagram for the model. The dashed line indicates a second order phase transition and the solid line a first order one. The point indicates where the second order phase transition turns to first order and corresponds to $(T,\mu)\approx$(105 MeV, 150 MeV).}
\label{Tmu}
\end{figure}

As shown in Fig. \ref{Tmu} the second order phase transition turns into a first order one around $(T,\mu)\approx$(105, 150 MeV). The diagram has the usual form and exhibits the behaviour one would expect for the model.

\section{Conclusions}

We have developed the real time formalism for a nNJL model in the chiral limit 
with a fractional Lorentzian regulator obtained in recent nonlocal NJL models 
which try to match lattice results on the quark propagator. 
Instead of the two regulators used in such model, here we only consider one of them   that produces corrections to the mass,  neglecting the regulator that produces a nontrivial  wave function renormalization term in the infrared sector. 
We obtained all the different quasiparticles masses and decay widths and their 
thermal behavior, in order to decide which of them will be relevant near the chiral phase 
transition. 
Due to the  pressence of the cut in the complex plane, the singularities are doubled. However, one of the main conclusions of this article is that only physical 
poles with a positive squared mass have to be considered.
 Not doing this, will lead 
to inconsistencies like a condensate that grows with $T$. So, the Wick 
rotation cannot be performed  in a simple way, and those unphysical terms must 
be removed by changing the path of integration shown in Fig. \ref{pathS} 
to obtain the appropriate spectral function.

Although similar analytic procedures can be used to compute the sum of the Matsubara frequencies, this direct treatment allows us to explore the system in different 
scenarios, explicitly including only the relevant quasiparticles 
that participate in the dynamics of the system, depending on their thermal 
behavior. An extension to the case where the chiral symmetry is explicitly broken can be 
obtained directly as we have shown. 
The reasonable values obtained for the critical temperature and 
critical chemical potential provide support for this procedure  compared with full model.

We would like to apply this technique to  the Keldish formalism and to construct an out of equilibrium effective model.

\section{Acknowledgments}

The authors would like to acknowledge support from FONDECYT under Grant No. 1130056. M.L. also acknowledge support from FONDECYT under Grant 1120770. F.M. would like to acknowledge support from CONICYT under Grant No. 21110577. The authors would like to thank R. Zamora for helpful discussions. F.M. would like to thank T. Cohen for a useful discussion. The authors thank D. Blaschke, S. Benic, and M. Buballa for a valuable correspondence.


\begin{thebibliography}{99}

\bibitem{Lattice1} M. Hanada, Y. Matsuo and N. Yamamoto, Phys. Rev. D. {\bf 86}, 074510 (2012).
\bibitem{Lattice2} J. Danzer, C. Gattringer, L. Liptak and M. Marinkovic, Phys. Lett. B. {\bf 682}, 240 (2009).
\bibitem{Jona} Y. Nambu and G. Jona-Lasinio, Phys. Rev. {\bf 122}, 345 (1961).
\bibitem{Lasinio} Y. Nambu and G. Jona-Lasinio, Phys. Rev. {\bf 124}, 246 (1961).
\bibitem{Loewe1} M. Loewe, Jorge Ruiz A. and J. C. Rojas, Phys. Rev. D {\bf 78}, 096007 (2008).
\bibitem{Buballa} M. Buballa, Phys. Rep. {\bf 407}, 205 (2005).
\bibitem{Klevansky} S.P. Klevansky, Rev. Mod. Phys. {\bf 64}, 649 (1992).
\bibitem{Blaschke1} A. E. Radzhabov, D. Blaschke, M. Buballa and M. K. Volkov, Phys. Rev. D. {\bf 83}, 116004 (2011).
\bibitem{Blaschke2} D. Blaschke, P.Costa and Yu. L. Kalinovsky, Phys. Rev. D. {\bf 85}, 034005 (2012). 
\bibitem{nNJL1} R. D. Bowler and M. C. Birse, Nucl. Phys. {\bf A582}, 655 (1995).
\bibitem{nNJL2} R. S. Plant and M. C. Birse, Nucl. Phys. {\bf A628}, 607 (1998).
\bibitem{Michal1} M. Praszalowicz and A. Rostoworowski, Phys. Rev. D {\bf 64}, 074003 (2001).
\bibitem{Scoccola6} D. G\'omez Dumm and N. N. Scoccola, Phys. Rev. D {\bf 65}, 074021 (2002).
\bibitem{Scoccola1} S. Noguera and N. N. Scoccola, Phys. Rev. D. {\bf 78} 114002 (2008).
\bibitem{Loewe2} M. Loewe, P. Morales and C. Villavicencio, Phys. Rev. D {\bf 83}, 096005 (2011).
\bibitem{LeBellac} M. Le Bellac, {\em Thermal Field Theory} (Cambridge University Press, Cambridge, England, 1996).
\bibitem{Das} A. Das, {\em Finite Temperature Field Theory} (Worls Scientific, Singapore, 1997).
\bibitem{Buballa2} S. Beni\'c, D. Blaschke and M. Buballa, Phys. Rev. D. {\bf 86} 074002 (2012).
\bibitem{Keldysh1} S. Onoda, N. Sugimoto and N. Nagaosa, Prog. Theor. Phys. {\bf 116} 61 (2006).
\bibitem{Keldysh2} A. Kamenev and A. Andreev, Phys. Rev. B. {\bf 60}, 2218 (1999).
\bibitem{Keldysh3} C. Chamon, A. Ludwig and C. Nayak, Phys. Rev. B. {\bf 60}, 2239 (1999).
\bibitem{Scoccola5} S. Noguera and N. N. Scoccola, Phys. Rev. D {\bf 78}, 114002 (2008).
\bibitem{Kapusta} J. I. Kapusta, {\em Finite-Temperature Field Theory}, (Cambridge University Press, Cambridge, England, 1989).
\bibitem{Ojima1} I. Ojima, Annals Phys. {\bf 137}, 1 (1981).
\bibitem{Ojima2} H. Matsumoto, I. Ojima, and H. Umezawam, Annals Phys. {\bf 152}, 348 (1984).
\bibitem{Kobes} R. L. Kobes, G. W. Semenoff, and N. Weiss, Z. Phys. C. {\bf 29}, 371 (1985).
\bibitem{Landsman} N. P. Landsman, and C. G. van Weert, Phys. Rept. {\bf 145}, 141 (1987).
\bibitem{Scoccola2}  D. G\'omez Dumm, A. G. Grunfeld, and N. N. Scoccola, Phys. Rev. D. {\bf 74}, 054026 (2006). 
\bibitem{Scoccola3} V. Pagura, D. G\'omez Dumm, and N. N. Scoccola, Phys. Rev. D. {\bf 87}, 014027 (2013).
\bibitem{Scoccola4}  J.P. Carlomagno, D. Gomez Dumm, and N. N. Scoccola,  arXiv:1305.2969. 
\bibitem{Scoccola20} G.A. Contrera, D. G\'omez-Dumm and N.N. Scoccola, Phys.Rev. D {\bf 81},  054005 (2010).
\bibitem{Ratti} S. Roessner, C. Ratti, and W. Weise, Phys.Rev. D {\bf 75}, 034007 (2007).   
\bibitem{Blaschke3} S. Beni\'c, D. Blaschke, G. A. Contrera and F. Horvati\'c, arXiv:1306.0588.



\end{thebibliography}
\end{document}